\documentclass{revtex4}
\usepackage{amsmath}
\usepackage{amsthm}
\usepackage{epsfig}
\usepackage{graphics}
\usepackage{graphicx}
\usepackage{dcolumn}
\usepackage{bm}
\usepackage{multirow}
\usepackage{natbib}
\usepackage{lscape}
\usepackage{pdflscape}
\begin{document}

\title{An infrared probe of the insulator-to-metal transition in Ga$_{1-x}$Mn$_x$As and Ga$_{1-x}$Be$_x$As}
\author{B. C. Chapler,$^{1,\ast}$ R. C. Myers,$^2$ S. Mack,$^3$ A. Frenzel,$^1$ B. C. Pursley,$^1$ K. S. Burch,$^4$ E. J. Singley,$^5$ A. M. Dattelbaum,$^6$ N. Samarth,$^7$ D. D. Awschalom,$^3$ and D. N. Basov$^1$ }

\affiliation
{$^{1}$Physics Department, University of California-San Diego, La Jolla, California 92093, USA \\
$^{2}$Department of Materials Science and Engineering, Ohio State University, Columbus, Ohio 43210, USA \\
$^{3}$Center for Spintronics and Quantum Computation, University of California-Santa Barbara, California 93106, USA \\
$^{4}$Department of Physics \& Institute for Optical Sciences, University of Toronto, Toronto, Ontario, Canada M5S 1A7 \\
$^{5}$Department of Physics, California State University-East Bay, Hayward, California 94542, USA\\
$^{6}$Los Alamos National Laboratory, Los Alamos, New Mexico 87545, USA \\
$^{7}$Department of Physics, The Pennsylvania State University, University Park, Pennsylvania 16802, USA
}

\date{\today}


\begin{abstract} 
We report infrared studies of the insulator-to-metal transition (IMT) in GaAs doped with either magnetic (Mn) or non-magnetic acceptors (Be). We observe a resonance with a natural assignment to impurity states in the insulating regime of Ga$_{1-x}$Mn$_x$As, which persists across the IMT to the highest doping (16\%). Beyond the IMT boundary, behavior combining insulating and metallic trends also persists to the highest Mn doping. Be doped samples however, display conventional metallicity just above the critical IMT concentration, with features indicative of transport within the host valence band.  
\end{abstract}

\maketitle
 
The insulator-to-metal transition (IMT) becomes exceptionally complex when magnetism is involved, as proven in materials such as mixed-valence manganites, rare-earth chalcogenides, and Mn-doped III-V compounds~\cite{Coey1999, Sato2010}. In all these systems, the electronic and magnetic properties are typically interconnected, creating an entising challenge to understand how magnetism affects the IMT physics. A promising route to isolate differences attributable to the presence of magnetism on the IMT physics is to investigate either magnetic or non-magnetic dopants in the same host. $p$-doped GaAs is well suited for the task since metallicity in this material can be initiated by non-magnetic (Zn, Be, C) and magnetic (Mn) acceptors. Infrared (IR) experiments reported here for Ga$_{1-x}$Be$_{x}$As and Ga$_{1-x}$Mn$_{x}$As monitor the charge dynamics and electronic structure in the course of the IMT. Our results establish that the onset of conduction in magnetically doped GaAs is distinct from genuine metallic behavior due to extended states in the host valence band (VB), as seen in moderately doped Ga$_{1-x}$Be$_{x}$As. Moreover, we observe a co-existence of ``metallic'' and ``insulating'' trends over a broad range of Mn concentrations (1-16\%), underscoring the unconventional nature of Mn-doped GaAs beyond the IMT boundary.

\begin{table}[t]
\centering
\includegraphics[width=86mm]{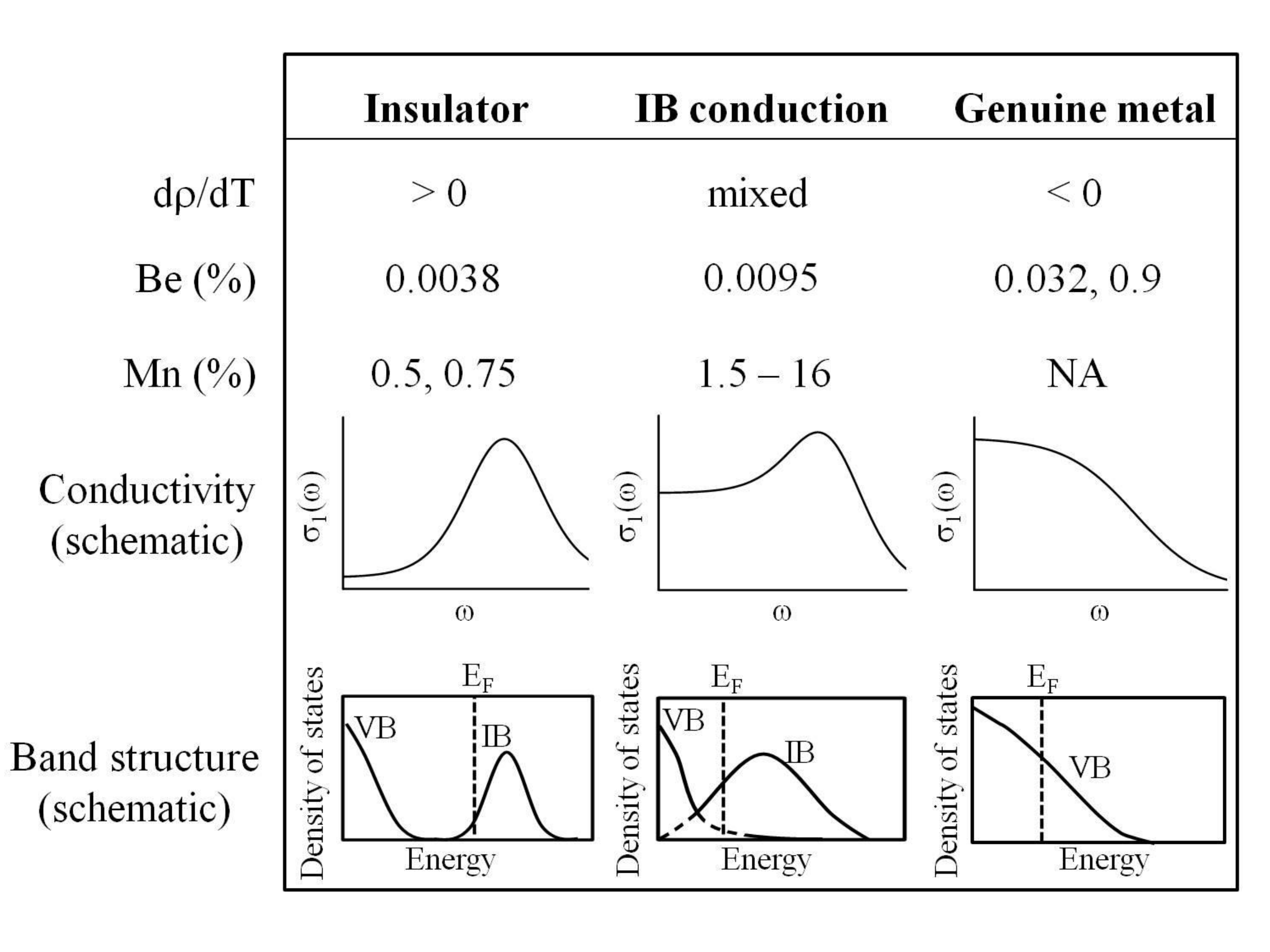}
\caption{Summary of the three transport regimes uncovered for $p$-doped GaAs. Conductivity schematics in the insulating and impurity band conduction regimes highlight the resonance associated with valance band to impurity band transitions. This is contrasted by the condutivity schematic in the genuinely metallic regime that highlights the Drude response due to free-carriers. We only list Mn and Be dopings reported in this study.
}
\label{table1}
\end{table}

The Ga$_{1-x}$Mn$_{x}$As samples were prepared using a non-rotated, low-temperature, molecular-beam-epitaxy (MBE) technique, which has been shown to minimize the formation of compensating defects~\cite{Myers2006, Mack2008}. Several samples were also subjected to post-growth annealing in attempt to further reduce/eliminate defects. The IMT in Ga$_{1-x}$Be$_{x}$As occurs at much lower dopant concentrations due to the much lower acceptor level ($E_{\mathrm{Be}}$=28 meV, $E_{\mathrm{Mn}}$=112 meV~\cite{Nagai2005}), thus samples near the Ga$_{1-x}$Be$_{x}$As IMT were grown using conventional equilibrium MBE. Importantly, the most metallic Ga$_{1-x}$Be$_{x}$As sample, Ga$_{0.991}$Be$_{0.009}$As, was grown under the same non-rotated, low-temperature conditions as the Ga$_{1-x}$Mn$_{x}$As samples. We can therefore expect nominally similar levels of disorder and charge density as that of Ga$_{1-x}$Mn$_{x}$As samples with $x$$\sim$0.01. Details on sample growth can be found in the supplementary material.

\begin{figure}[t]
\centering
\includegraphics[width=86mm]{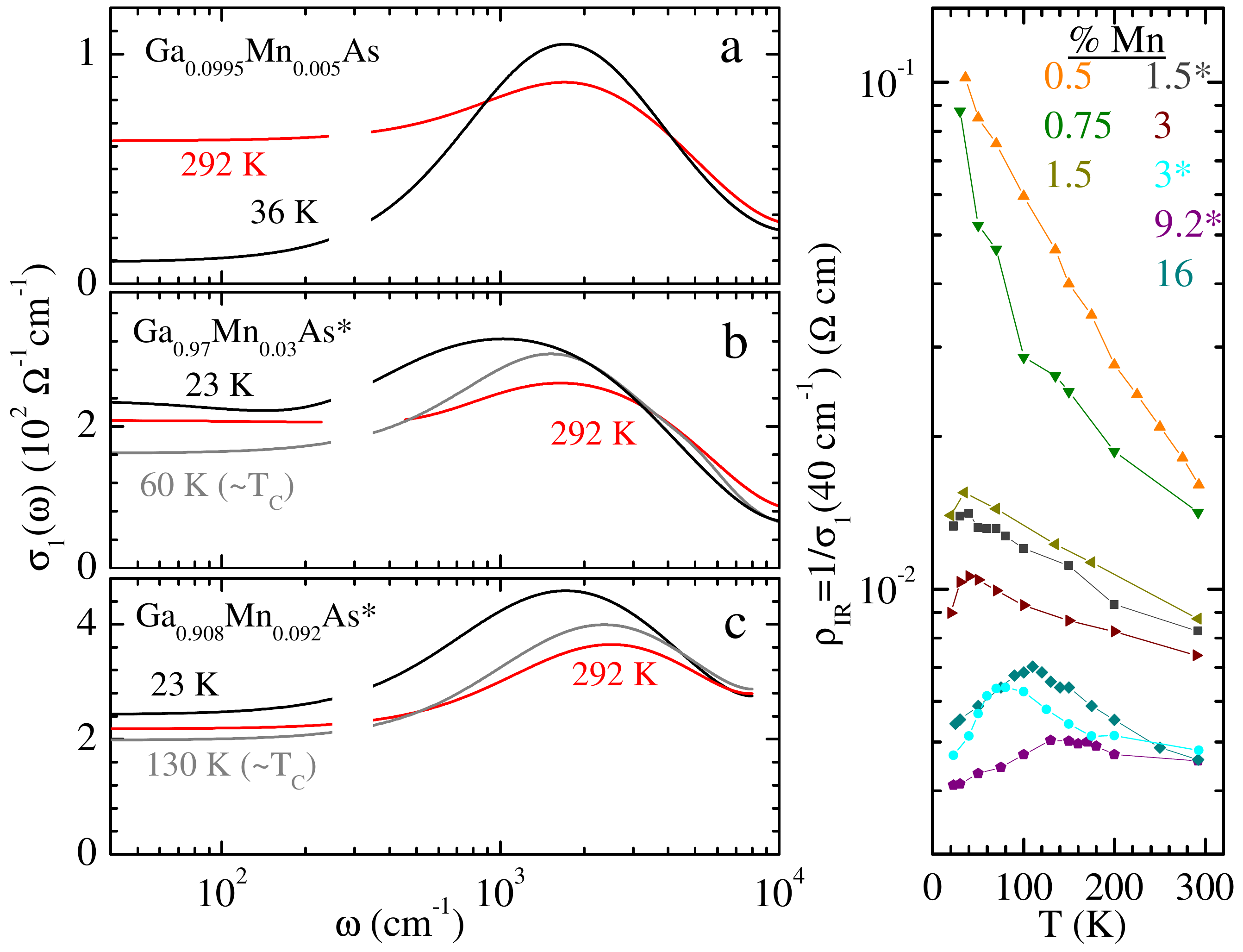}
\caption{Characteristic spectra of the Mn-doped samples in the insulating (a) and IB conduction (b and c) regimes are exemplified by the Ga$_{0.995}$Mn$_{0.005}$As, Ga$_{0.97}$Mn$_{0.03}$As*, Ga$_{0.908}$Mn$_{0.092}$As* samples, respectively. The break in the spectra is due to a phonon in GaAs which does not allow $\sigma_1(\omega$) to be extracted over this region.  d) The temperature dependence of $\rho_{\mathrm{IR}}$ for all Mn-doped samples in the study.
}
\label{fig3}
\end{figure}

Through a combination of transmission and ellipsometric measurements, we have probed the optical properties of the samples over a broad range of frequencies, spanning $\sim$40--40,000 cm$^{-1}$ (experimental details can be found in the supplemental material). Here we primarily focus on the real (or dissipative) part of the optical conductivity ($\sigma_1(\omega$)) from THz to just below the GaAs band gap energy ($\sim$40--10,000 cm$^{-1}$). Through this probe, signatures of three distinct electronic transport regimes in \emph{p}-doped GaAs are revealed (see Table~\ref{table1}): a regime of insulating behavior; a regime of genuine metallic behavior (i.e. no signs of thermal activation) due to extended states in a partially unoccupied VB; and an unusual intermediate regime of conduction, exhibiting elements of both insulating and metallic transport. We refer to this latter regime as impurity band (IB) conduction, with this term used to describe the persistence of a resonance associated with VB to IB transitions beyond the onset of conductivity.  

We begin by discussing the insulating regime of Ga$_{1-x}$Mn$_x$As. Fig.~\ref{fig3}a depicts the spectra for a paramagnetic Ga$_{0.995}$Mn$_{0.005}$As sample. The dominant feature of these data is a broad mid-IR resonance, with peak frequency ($\omega_0$) near 1700 cm$^{-1}$. Room-temperature data show substantial spectral weight at far-IR frequencies (Fig.~\ref{fig3}a). However, far-IR weight is transferred to the mid-IR resonance as the sample is cooled, and ``freezes out'' at low temperature. Elimination of far-IR spectral weight unambiguously reveals the thermally activated nature of the electronic transport in this dilute regime. In Fig.~\ref{fig3}d, we plot the temperature dependence of the ``infrared resistivity'' ($\rho_{\mathrm{IR}}$=1/$\sigma_1$(40 cm$^{-1})$). The data in dilute Mn-doped samples ($x$=0.005, 0.0075) show the systematic increase in $\rho_{\mathrm{IR}}$ expected in the case of thermally activated transport, in support of the conclusion of variable-range hopping inferred from direct resistivity measurements\cite{Sheu2007}. The existence of a narrow impurity band is not in dispute in dilutely doped insulating samples~\cite{Moriya2003, Jungwirth2007}, thus the observed mid-IR resonance in the vicinity of $E_{\mathrm{Mn}}$ in insulating samples can be naturally assigned to VB to IB transitions. We note although this is the first detailed report of a mid-IR resonance in the insulating state, numerous probes report evidence of an IB in moderately doped Ga$_{1-x}$Mn$_x$As\cite{Okabayashi2001, Singley2002, Burch2006, Kojima2007, Ando2008, Ohya2010a, Mayer2010, Sapega2005, Rokhinson2007}. 

Fig.~\ref{fig3}b and c show the spectra of the ferromagnetic Ga$_{0.97}$Mn$_{0.03}$As* and Ga$_{0.908}$Mn$_{0.092}$As* samples (* denotes annealed), the highest conducting Ga$_{1-x}$Mn$_{x}$As samples in this study. The spectra of both of these samples are characteristic of the IB conduction regime. The dominant feature of the spectra is a mid-IR resonance, similar in shape and $\omega_0$ to the insulating sample. These samples additionally show a finite $\sigma_{\mathrm{DC}}$ in the limit of $\omega$, T$\rightarrow$0: a condition often associated with the onset of metallicity. Nevertheless, a substantial fraction of the far-IR spectral weight still reveals activated behavior. Specifically, as the temperature is lowered from 300 K to near T$_C$, the far-IR spectral weight is suppressed and transferred to the mid-IR resonance, similar to data in Fig.~\ref{fig3}.

The onset of ferromagnetism radically alters the temperature dependence of $\sigma$$_1$($\omega$). Below T$_C$, we observe the reversal of the activated character (black spectra in the panels b and c), as the far-IR spectral weight, as well as the mid-IR resonance are enhanced. Earlier data have established a correlation between the enhancement of the spectral weight and the development of the magnetization~\cite{Singley2002}. Such enhancement serves as an unmistakable signature of the deep bond between magnetism and carrier dynamics in this class of carrier-mediated ferromagnets and other magnetic materials\cite{Coey1999, Okimoto1995, Hirakawa2001, Basov2011}. Comparison with dilute samples show the $\rho_{\mathrm{IR}}$ data still display (weakly) activated transport above T$_C$. Peaks in $\rho_{\mathrm{IR}}$ near T$_C$ mark the reversal of activated trends.

Our data show the onset of conduction in Ga$_{1-x}$Mn$_x$As occurs at a doping level 0.0075$<x\leq$0.015, in excellent agreement with the Mott-criterion ($p_c^{1/3}$=2.78$a_H$, where $p_c$ is the critical acceptor concentration and $a_H$ is the effective Bohr radius of acceptor holes), corresponding to $x$=0.0109 for Ga$_{1-x}$Mn$_{x}$As (assuming 1 hole/Mn and no compensation)~\cite{Nagai2005}. This agreement attests to a low degree of compensation in our samples, at least at doping regimes below a few atomic percent. We further note the IB conduction regime persists over an order of magnitude in dopant concentration, corresponding to at least $x$=0.015--0.16 in Ga$_{1-x}$Mn$_{x}$As, as well as in heavily doped annealed samples (Ga$_{0.908}$Mn$_{0.092}$As*). No Mn-doped samples in this study were found to exhibit genuinely metallic behavior, which we report and describe below for Ga$_{1-x}$Be$_x$As.


\begin{figure}[t]
\centering
\includegraphics[width=86mm]{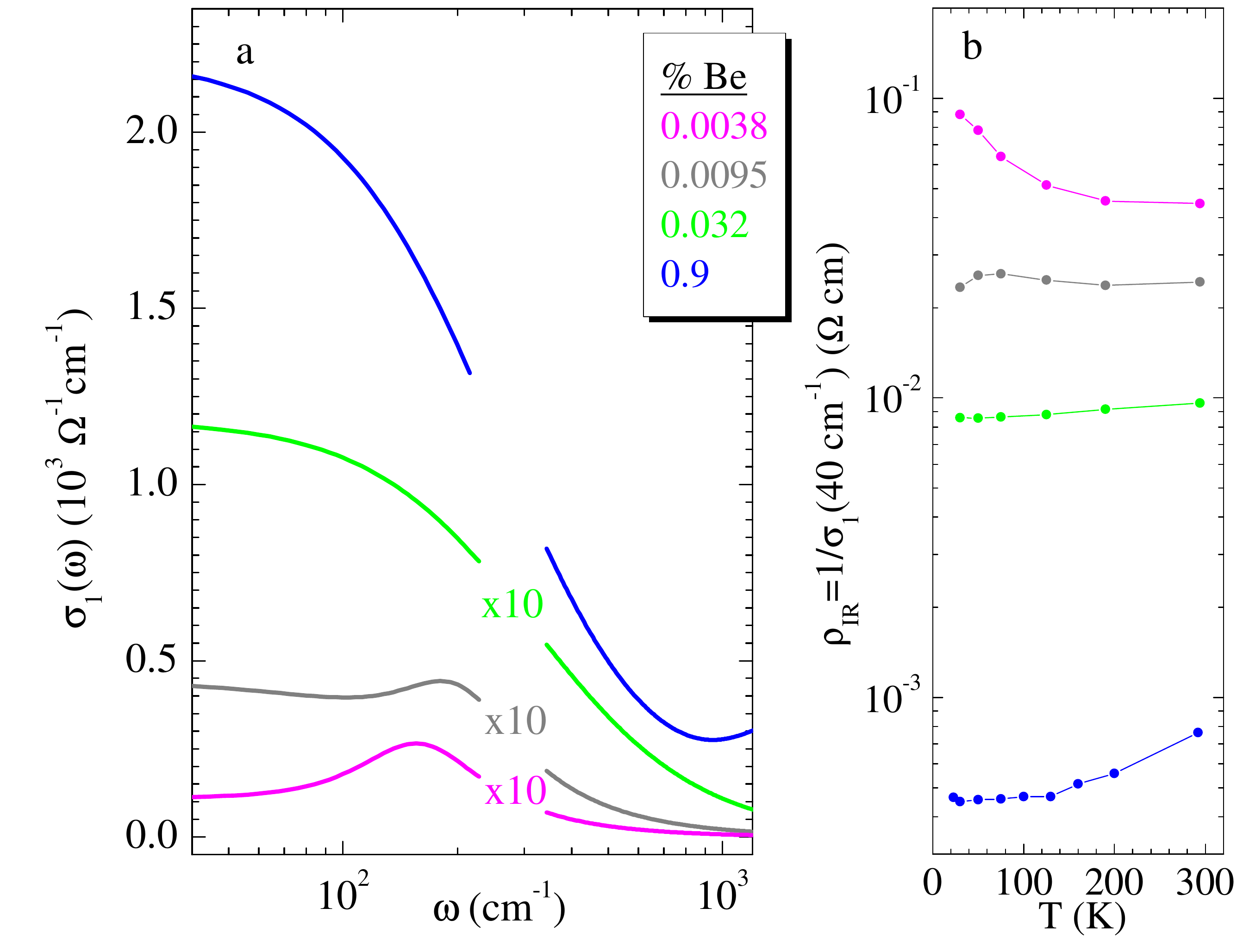}
\caption{a) Low temperature ($\sim$25 K) spectra for the Ga$_{1-x}$Be$_x$As samples. All spectra but that for the $x$=0.009 sample are multiplied by 10 to show on the same scale. b) Temperature dependence of $\rho_{\mathrm{IR}}$ for all Be-doped samples.}
\label{gabeasimt}
\end{figure}

Data in Fig.~\ref{gabeasimt} display the IMT in Ga$_{1-x}$Be$_{x}$As. The $x$=3.8$\times$10$^{-5}$ and $x$=9.5$\times$10$^{-5}$ Be-doped samples (pink and grey curves in Fig.~\ref{gabeasimt}) reveal significant low-frequency conductivity at room temperature. However, the conductivity of the $x$=3.8$\times$10$^{-5}$ Be-doped sample is frozen out at low-temperatures, analogous to the insulating Ga$_{1-x}$Mn$_x$As samples. Also similar to Ga$_{1-x}$Mn$_x$As, both $x$=3.8$\times$10$^{-5}$ and $x$=9.5$\times$10$^{-5}$ Be-doped samples show a resonance centered in the vicinity of the $E_{\mathrm{Be}}$. Such a feature near the acceptor level reinforces a VB to dopant-induced IB interpretation in both Mn- and Be-doped GaAs. The $\rho_{\mathrm{IR}}$ data for the $x$=9.5$\times$10$^{-5}$ sample appear to show a finite $\sigma_{\mathrm{DC}}$ in the limit of $\omega$, T $\rightarrow$0, implying this sample may be past the onset of conduction. Thus an intermediate IB conduction regime could be a generic feature of $p$-doped GaAs (and potentially many other doped semiconductors) near the IMT~\cite{Romero1990, Gaymann1995}.

Moving to the $x$=3.2$\times$10$^{-4}$ and $x$=0.009 Be-doped samples (light-green and blue curves in Fig.~\ref{gabeasimt}), we see spectra qualitativley different from those observed in samples either in the insulating or IB conduction regimes. Here conductivity data are dominated by a pronounced Drude peak (Lorentzian centered at $\omega$=0), characteristic of delocalized carriers in a metal. Insights into the nature of the metallic state in Ga$_{1-x}$Be$_{x}$As is revealed by applying the partial sum rule to the conductivity:

\begin{equation}
{\int_{0}^{\omega_c}\sigma_{1}(\omega)d\omega}={\frac{\pi pe^{2}}{2m_{\mathrm{opt}}}}.
\label{drude}
\end{equation}

\noindent This analysis yields the effective optical mass ($m_{\mathrm{opt}}$) provided the carrier density $p$ is known. Using an integration cut-off of $\omega_c$=6450 cm$^{-1}$~\cite{cutoff}, we obtain $m_{\mathrm{opt}}=0.28 m_e$ and 0.29$m_e$ ($m_e$ is the electron mass) for the $x$=3.2$\times$10$^{-4}$ and $x$=0.009 Be-doped samples, respectivley. The extracted $m_{\mathrm{opt}}$ is in excellent agreement with calculations of the IR spectra of metallic $p$-doped GaAs with a partially unoccupied VB, which place $m_{\mathrm{opt}}$ between 0.25$m_e$ and 0.29$m_e$\cite{Sinova2002}. These results, along with the metallic temperature dependence (green and blue curves in Fig.~\ref{gabeasimt}b), give strong evidence of transport due to light quasiparticles in a partially unoccupied VB. Note, the Mott-criterion for Be-doped GaAs predicts a critical dopant concentration of $x$=2.7$\times$10$^{-4}$~\cite{Nagai2005}. Thus our data indicate genuinely metallic behavior is achieved even for a 20\% increase in dopant density beyond the Mott critical concentration. This stands in stark contrast to the case of Mn-doped samples.

\begin{figure}[t]
\centering
\includegraphics[width=86mm]{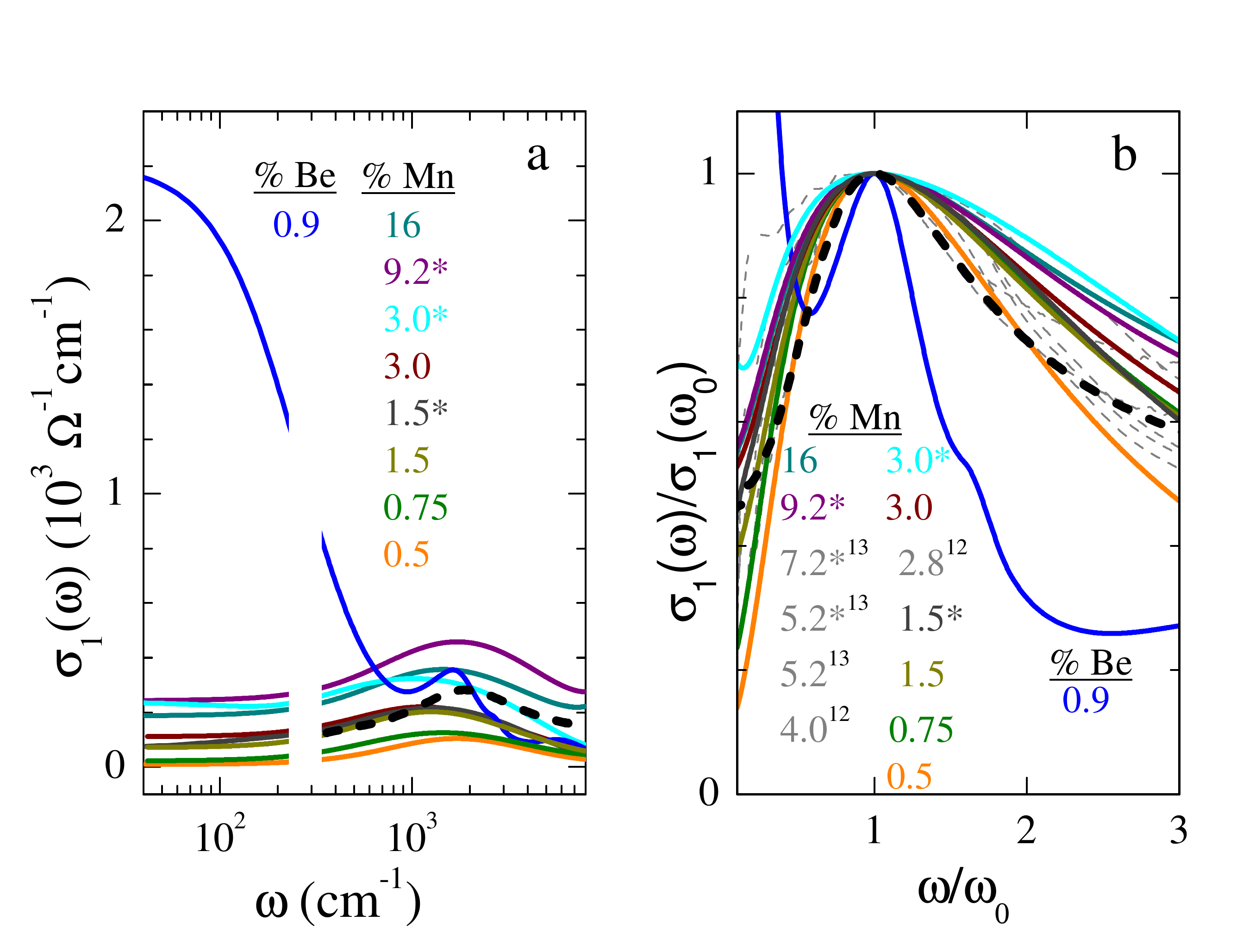}
\caption{a) Direct comparison of the conductivity spectra displayed for the lowest temperatures measured (20-36 K) of the $x$=0.009 Be-doped film and all Mn-doped films studied. b) The mid-IR resonance in the two classes of materials is directly compared by normalizing spectra by the peak in $\sigma$$_{1}$($\omega$) at $\omega_0$ along the $y$-axis, and by $\omega_0$ along the $x$-axis. Also shown in both panels is calculated $\sigma$$_{1}$($\omega$) according to the quantum defect method (black dashed)~\cite{Kojima2007}. Additional Ga$_{1-x}$Mn$_x$As data are reported from Burch $\it{et}$ al.\cite{Burch2006} and Singley $\it{et}$ al.\cite{Singley2002}.
}
\label{fig2}
\end{figure} 

To further highlight the differences between magnetic and non-magnetic dopants in GaAs, we plot our results for the Ga$_{1-x}$Mn$_x$As samples, as well as the Ga$_{0.991}$Be$_{0.009}$As sample in Fig.~\ref{fig2}a. Fig.~\ref{fig2}b shows scaled conductivity for these samples, as well as earlier IR data for Ga$_{1-x}$Mn$_x$As, normalized by $\sigma$$_{1}$($\omega$) at $\omega_0$ along the $y$-axis, and by $\omega_0$ along the $x$-axis. In both Mn- and Be-doped samples we observe a mid-IR resonance, however a result that stands out in both panels is the difference in lineshape of this resonance in genuinely metallic Ga$_{1-x}$Be$_x$As and that in the Mn-doped samples. The resonance seen in Ga$_{0.991}$Be$_{0.009}$As is much narrower and reveals a two-peak structure. The second peak appears as a shoulder on the main peak, and is most clearly seen near $\omega/\omega_0$=1.5 in Fig.~\ref{fig2}b. The frequency position of the two-peak structure is an order of magnitude higher than $E_{\mathrm{Be}}$, however it is near the energy expected for intra-VB transitions, in accord with $E_F$ located deep within the VB. The main peak is thus attributed light-hole band (LH) to heavy-hole band (HH) excitations~\cite{Songprakob2002}, while the shoulder is due to excitations from the split-off band (SO)\cite{Braunstein1962}.

In contrast to our observations in the Ga$_{0.991}$Be$_{0.009}$As film, all the Mn-doped samples (both in the insulating regime and those past the onset of conductivity) show very broad, structureless absorption. Furthermore, upon scaling of these data (Fig.~\ref{fig2}b), the Mn-doped samples reveal a nearly identical lineshape, barring small non-monotonic differences in the width mid-IR peak resonance. The similar lineshape is indicative of a similar origin of the mid-IR resonance in both insulating and conducting samples (VB to IB optical excitations). The scaled Ga$_{1-x}$Mn$_x$As data include early IR results, with different growth procedures (rotated and non-rotated, post growth annealing, etc.) that imply variable levels of disorder, yet the qualitative features reveal only minimal dependence on the degree of disorder. These facts allow us to conclude that disorder is not playing a major role in the infrared response of Ga$_{1-x}$Mn$_{x}$As. Another result evident from Fig.~\ref{fig2} is the prominent Drude peak observed only in Ga$_{1-x}$Be$_x$As samples. Much weaker low-energy spectral weight is exhibited in Mn-doped samples, even those subjected to low-temperature annealing known to enhance the carrier density\cite{Yu2002} (further discussion of low-$\omega$ analysis of Be and Mn-doped samples can be found in supplemental materials). All these findings demonstrate the distinct nature of transport in films with magnetic Mn-dopants.   

Simple models, such as that calculated according to the quantum defect method for band-to-acceptor transitions (black dashed Fig.~\ref{fig2})~\cite{Kojima2007} (see supplementary material for model details), as well as more complex calculations~\cite{Moca2009a, Bouzerar2010a} been shown to successfully capture key aspects of the mid-IR resonance in Ga$_{1-x}$Mn$_x$As. However, other explanations of the mid-IR resonance have beeen proposed~\cite{Jungwirth2007}, and the microscopic justification of such ``impurity band models'' have been called into question~\cite{Masek2010}. Our systematic study showing the mid-IR resonance exists in the insulating regime, and persists with the same lineform across the IMT up to the highest attainable doping, conclusively demonstrates impurity states dominate electronic dynamics in Ga$_{1-x}$Mn$_{x}$As, even in highly conductive samples. 

Aspects of coexistence of the IB and metallicity are perhaps understandable very near the IMT, as in the case of the $x$=9.5$\times$10$^{-5}$ Be-doped sample. However our results show that in the magnetically doped films, signitures of IB states remain, along with the co-existence of metallic and insulating trends, at least an order of magnitude in dopant concentration beyond that associated with the onset of conduction. Theororetical work has directly linked exchange coupling between Mn-ions and valence holes to the persistence of the IB in Ga$_{1-x}$Mn$_x$As~\cite{Nili2010a, Chattopadhyay2001}, further underscoring the pivotal role of magnetism in the electronic structure and optical phenomena of Ga$_{1-x}$Mn$_x$As. Because trivial clustering of multiple phases has been ruled out in Ga$_{1-x}$Mn$_x$As~\cite{Dunsiger2010, Richardella2010a}, the duality between metallic and insulating trends appear to be an intrinsic attribute of the IMT in magnetically $p$-doped GaAs~\cite{Sapega2009}. In view of the above duality, it may not be surprising that some properties of Ga$_{1-x}$Mn$_x$As can be explained from the standpoint of Bloch-like states~\cite{Jungwirth2006, Dietl2000, Masek2010}. However, many salient features of Ga$_{1-x}$Mn$_x$As cannot be understood in the context of conventional VB conduction, such as the spectral features and activated trends seen in Fig.~\ref{fig3}, and require an emphasis on localization~\cite{Moca2009a, Bouzerar2010b, Bouzerar2010a}. We speculate electronic correlations are likely to be vitally important in a conducting system at the borderline of localization, and note earmarks of such electron-electron effects have been identified in Ga$_{1-x}$Mn$_x$As tunneling spectra~\cite{Richardella2010a}. Though disorder plays a role, our experiments show that disorder alone is not sufficient to explain the radical differences between Be-doped and Mn-doped samples. 

Work at UCSD is supported by the Office of Naval Research. Work at UCSB is supported by the Office of Naval Research and the National Science Foundation.

\section{Supplementary Material}

\subsection{Materials and Methods}

All films in this study were prepared at UCSB using molecular-beam-epitaxial (MBE) growth, on insulating (001) GaAs substrates. Since this study involved several doping regimes, some of the details of the growth vary between samples. All of the Ga$_{1-x}$Mn$_x$As films studied were prepared following the non-rotated, low-temperature growth techniques reported in Ref.\cite{Myers2006, Mack2008} (see \ref{growth}). Specific details on growth properties of all samples can be found in Table~\ref{props}. The $x$=0.015, 0.03, and 0.092 samples were also subjected to low-temperature annealing, which has been shown to remove Mn-interstitials in Ga$_{1-x}$Mn$_x$As, increasing the carrier density $p$ and T$_C$, as well as improving other sample properties~\cite{Edmonds2004}. Note the $x$=0.015 and 0.015* Ga$_{1-x}$Mn$_x$As films are the same sample pre- and post-annealing, as well as for the $x$=0.03 and 0.03* Ga$_{1-x}$Mn$_x$As films (* indicates the sample was annealed). Ga$_{1-x}$Mn$_x$As samples of \emph{x}=0.005, 0.0075 were found to be paramagnetic (PM) at all locations along the films. The other more heavily doped Ga$_{1-x}$Mn$_x$As films were found to be ferromangetic (FM), with T$_C$ in the region of minimized compensating defects listed in Table~\ref{props}. In the $x$=0.015* and 0.03* Ga$_{1-x}$Mn$_x$As samples, T$_C$ after annealing was determined by the temperature of maximum low-frequency resistivity ($\rho_{\mathrm{IR}}$=1/$\sigma_1$(40 cm$^{-1}$)). For all other samples, including $x$=0.015 and $x$=0.03 before annealing, T$_C$ was measured in a SQUID magnetometer. The Ga$_{0.991}$Be$_{0.009}$As was grown to a nominal thickness of $t$=100 nm, also using the non-rotated, low-temperature technique at $250\,^{\circ}{\rm C}$. The Ga$_{1-x}$Be$_{x}$As samples of $x$=3.8$\times$10$^{-5}$, 9.5$\times$10$^{-5}$, and 3.2$\times$10$^{-4}$ do not require low-temperature conditions during growth because of their relatively low doping concentrations. The Mn dopant concentration was determined by growth rate calibrations of MnAs and GaAs RHEED oscillations, while the concentration in Be-doped samples was determined from room temperature Hall-effect measurements. Deviations between $p$ extracted from Hall measurements and actual Be concentration are expected to be minor due to the small, and in some cases non-existent, activation energy in the Be-doped samples.

\begin{table*}
\centering
\begin{tabular}{c ||c | c | c | c | c} 
\hline 
dopant & concentration (\%) & nominal thickness (nm) & growth temperature ($^{\circ}{\rm C}$) & annealing temperature ($^{\circ}{\rm C}$) &  T$_C$ (K)\\ 
\hline\hline
\multirow{8}{*}{Mn} & 0.5 & 200 & 250  & - & $<$ 4  \\ 
 & 0.75 & 200 & 250  & - & $<$ 4  \\
 & 1.5 & 200 & 250  & - & 21  \\
 & 1.5* & 200 & 250  & 220 & 21  \\
 & 3.0 & 200 & 220  & - & 42  \\
 & 3.0* & 200 & 220  & 200 & 60  \\
 & 9.2* & 100 & 200  & 180 & 130  \\
 & 16 & 100 & 150  & - & 120  \\
\hline
\multirow{4}{*}{Be} & 0.0038 & 980 & 580  & - & - \\ 
 & 0.0095 & 980 & 580  & - & - \\
 & 0.032 & 980 & 580  & - & - \\
 & 0.9 & 100 & 250 & - & -  \\
\hline \hline
\end{tabular}
\caption{Sample growth properties.}
\label{props}
\end{table*}

All data presented in the manuscript on films grown the non-rotated technique were collected from d$\sim$0.5 mm spot where As-flux is tuned to maximize $p$. We therefore conclude that at this special location, compensation is minimized, as is disorder in general. This conjecture is supported through detailed transport studies of films prepared using similar growth protocol~\cite{Myers2006, Mack2008}. For each sample, this particular location was established through systematic studies of IR spectra along the wafer, and determined by the location of the largest integrated spectral weight (see \ref{gradient}). Furthermore, room temperature transport measurements have confirmed that the location of maximum $p$ corresponds also to the minimum resistivity, highest mobility, and in the case of FM samples, maximum T$_C~$\cite{Myers2006, Mack2008}.

\subsection{Non-rotated growth}
\label{growth}

Several films in the study (Ga$_{1-x}$Mn$_x$As: $x$=0.005, 0.0075, 0.0115, 0.03, 0.09, and 0.16; Ga$_{1-x}$Be$_x$As: $x$=0.009) were prepared using the non-rotated, low-temperature, MBE growth techniques reported in Ref.~\cite{Myers2006, Mack2008}. The low growth temperatures are required to suppress precipitation of secondary crystal phases, such as MnAs. However, the low growth temperatures also result in the formation of compensating defects. Of these defects, As-antisites (As$_{\mathrm{Ga}}$) are particularly problematic in that they are known to be double-donors (compensating two holes)\cite{Liu1995}, can reach concentrations on the order of 10$^{20}$ cm$^{-3}$ ($\sim$1\% of Ga sites)\cite{Missous1994}, and cannot be annealed out. The non-rotated growth method is a novel technique that systematically reduces the presence of As$_{\mathrm{Ga}}$.

The non-rotated growth utilizes a geometry in which the MBE system provides a continuous variation in the As:Ga ratio across the wafer. Control of the As$_{\mathrm{Ga}}$ content is achieved by the precise tuning of the As-flux, via the As:Ga gradient. Studies along the As:Ga gradient reveal an optimized location along the wafer where As$_{\mathrm{Ga}}$ has been minimized. This optimized location has been shown to correspond to maximum hole density $p$, maximum mobility, and minimum resistivity\cite{Myers2006}. In ferromagnetic samples, the optimized location also reveals maximized T$_C$ for given Mn concentrations, as well as improved magnetization, and magnetic hysteresis curves\cite{Myers2006, Mack2008}.

\subsection{Optical characterization}
\label{gradient}

The non-rotated technique provides a continuously varying As:Ga growth ratio, and thus yields a film with a gradient density of compensating As-antisite defects across the wafer. Local transport and local optical measurements in principle enable studies of \emph{p}-doped GaAs at the location where the density of these defects is minimized. In order to characterize the films in this study, extensive infrared (IR) spectroscopic measurements were done along the As:Ga gradient in samples grown using the non-rotated technique. The location of maximum $p$ can be determined directly from transmission measurements, as maximum $p$ corresponds to the maximum absorption. For this purpose we developed a broad-band (far-IR to near-ultraviolet) microscope compatible with the low-temperature (20 K) operation. The spatial resolution of the apparatus is below 1 mm which is appropriate for the series of films we investigated here. This assertion was validated through direct IR microscopy experiments with spatial resolution less than 100 $\mu$m.

\subsection{Extraction of optical conductivity}
\label{extract}

Transmission spectra are dependent on both components of the complex conductivity spectra $\sigma(\omega$)=$\sigma_1(\omega$)+i$\sigma_2(\omega)$. Importantly, these two components are not independent, but linked through the Kramers-Kronig (KK) relations~\cite{Wooten1972}. Adding complexity, in multi-layer samples the data will include contributions to the optical response from all layers. A convenient method for overcoming the complications of multi-layer systems, and extracting $\sigma(\omega$) for a single layer in a multi-layer sample, is via multi-oscillator modeling. In the case of the film/substrate system studied here, the substrate $\sigma(\omega$) is initially characterized, thus only $\sigma(\omega$) of the film must be extracted. The film layer is modeled using multiple oscillators, all chosen to ensure Kramers-Kronig consistency. 

\begin{figure}[t]
\centering
\includegraphics[width=86mm]{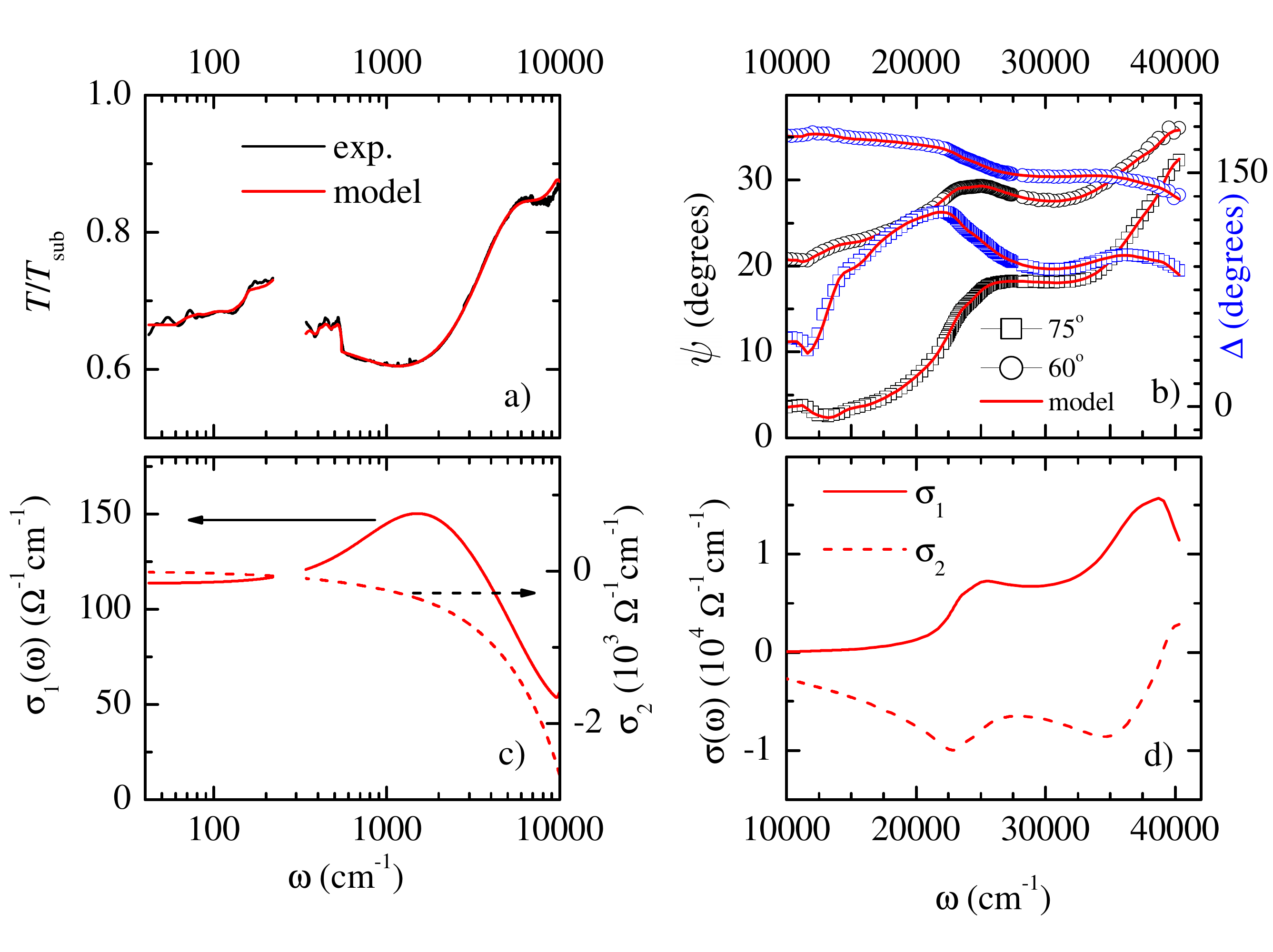}
\caption{Representative room-temperature transmission and ellipsometry data is shown with model fit to highlight $\sigma(\omega)$ extraction technique. All panels show data corresponding to the Ga$_{0.985}$Mn$_{0.015}$As sample. a) Experimental transmission spectrum normalized by transmission of the substrate, along with the multi-oscillator model fit. b) Ellipsometric angles $\psi$ and $\Delta$ (see Eq.~\ref{ellipseq}) measured at $75\,^{\circ}$ and $60\,^{\circ}$ angle of incidence is shown with the model fit. (c and d) Real ($\sigma_1(\omega)$) and imaginary ($\sigma_2(\omega)$) components of the $\sigma(\omega)$ spectrum resulting from the model are shown. Note red data in all four panels represent the same model, which provides a KK consistent fit of all experimental data across the entire frequency range. 
}
\label{everything}
\end{figure}

\begin{figure}[]
\centering
\includegraphics[width=86mm]{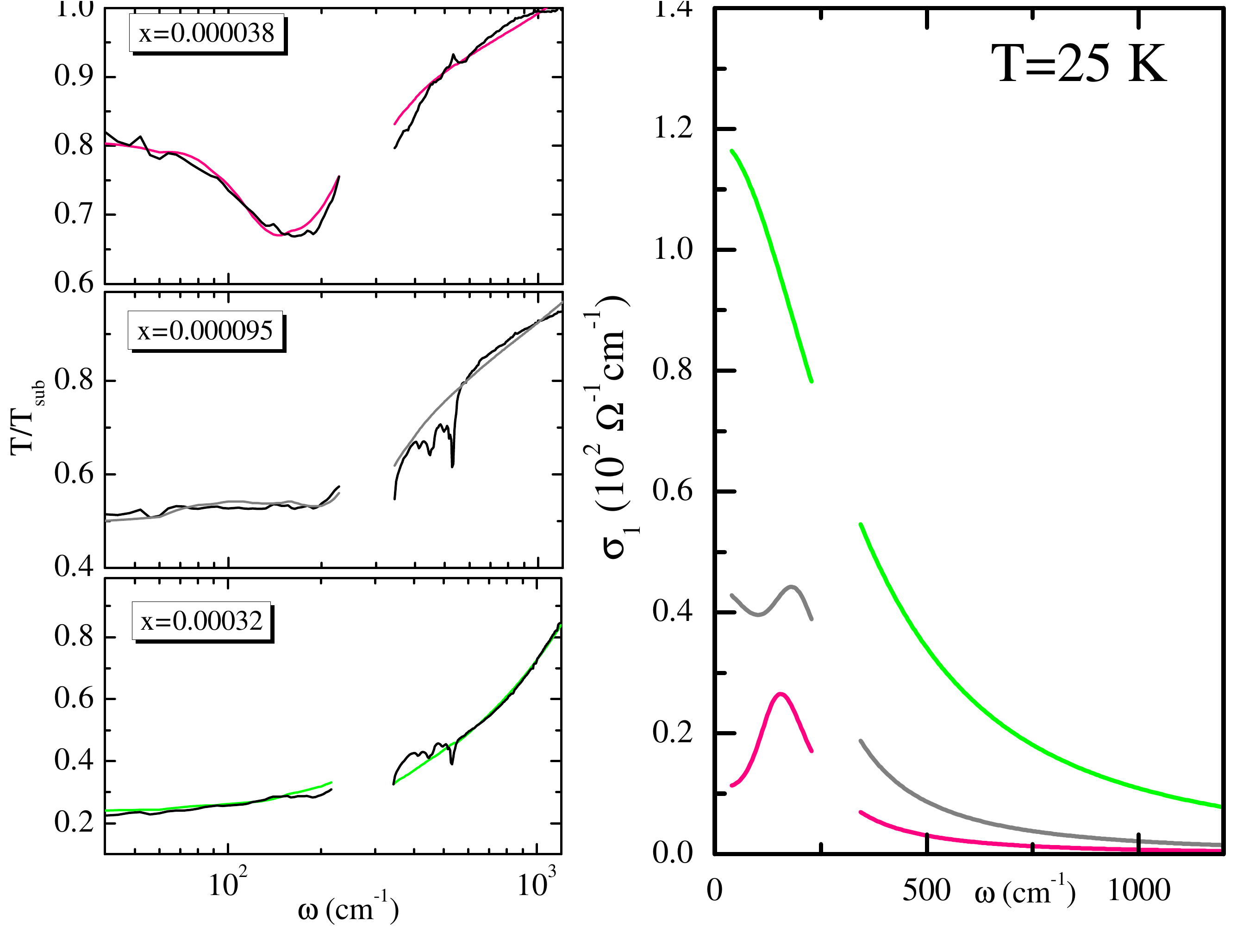}
\caption{(Left panels) Low temperature transmission of Ga$_{1-x}$Be$_x$As near the IMT normalized to the substrate transmission (black) along with model fits. (Right panel) Low temperature $\sigma_1(\omega)$ spectra extracted from the transmission data.
}
\label{beimt}
\end{figure}

Modelling of transmission data results in a KK consistent extraction of $\sigma(\omega)$, which reproduces the experimental data. However, $\sigma_2(\omega)$ in an intra-gap region is mainly determined by processes above the band gap. Therefore, accurate determination of the intra-gap $\sigma(\omega)$ must include experimental input from at frequencies above the gap. In the case of GaAs substrates, the samples become opaque at frequencies above the GaAs band gap ($\sim$12,000 cm$^{-1}$), placing a firm upper limit on the transmission data range. Thus, several films ($x$=0.005, 0.0075, 0.015, 0.03, 0.16 Mn and $x$=0.009 Be) were also measured at room temperature using micro-ellipsometry for 6,000 cm$^{-1}\leq \omega \leq$ 40,000 cm$^{-1}$, with spatial resolution d$\sim$150 $\mu$m. Ellipsometry measures the ratio of the complex reflectivity coefficient ($r$) of $s$- and $p$-polarized light. The experimental observables are the ellipsometric angles $\psi$ and $\Delta$ according to,

\begin{equation}
{\frac{r_p}{r_s}=\tan(\psi)e^{i\Delta}},
\label{ellipseq}
\end{equation}

\noindent where the subscript denotes the incident polarization. Ellipsometry data also have the additional advantage that they contain both magnitude and phase information. Thus, both components of $\sigma(\omega$) can be directly and uniquely extracted from the data over the entire ellipsometric frequency range via multi-oscillator modelling. One is therefore able, by simultaneous fitting of the model to the ellipsometry and transmission data, to produce KK consistent extraction of $\sigma(\omega$) over entire experimental range, with a high degree of confidence in the uniqueness and accuracy. Fig.~\ref{everything} displays representative plots illustrating the $\sigma(\omega)$ extraction procedure, showing experimental data and model fit, as well as the resulting $\sigma(\omega$) for the Ga$_{0.985}$Mn$_{0.015}$As sample. Fig.~\ref{beimt} shows the experimental transmission data and model fits of of the $x$=3.8$\times$10$^{-5}$, 9.5$\times$10$^{-5}$, and 3.2$\times$10$^{-4}$ samples, along with the $\sigma_1(\omega)$ curves for the corresponding models.

\subsection{Quantum Defect Model}
\label{defect}
The mid-IR resonance observed in the Ga$_{1-x}$Mn$_x$As was modeled using the quantum defect method~\cite{Bebb1967}. This model was first applied to absorption in semiconductors by Bebb $et$ al.~\cite{Bebb1969}, and later used to model absorption in magnetic semiconductors such as Cd$_{1-x}$Mn$_x$Te~\cite{Huber1987} and Ga$_{1-x}$Mn$_x$As~\cite{Kojima2007}. In this model, the cross section for photoionization transitions accociated with a given band is written as,
\begin{equation}
S(\hbar \omega)=\frac{4\pi \alpha_0}{3n(\hbar \omega)}[\frac{m^*}{m_0}]^2[\frac{E_{\mathrm{eff}}}{E_0}]^2\frac{2^{2\nu}{(\nu a^*)^2}f(y)}{y^{1/2}(1+y)^{\nu}}, \nonumber
\end{equation}
\begin{equation} 
f(y)=[\frac{\sin[(\nu+1)\tan^{-1}(y^{1/2})]}{y^{1/2}} - \frac{(\nu+1)\cos[(\nu+2)\tan^{-1}(y^{-1/2})]}{(1+y)^{1/2}}]^2,   
\label{model}
\end{equation}
\begin{equation}
y=\frac{\hbar \omega - E_i}{E_i}, \nonumber
\end{equation}
\noindent where $n$($\hbar \omega$) is the frequency dependent index of refraction, $\alpha_0$ the fine structure constant, $\nu$=($R^*$/E$_i$)$^{1/2}$=($e^4m^*$/2$\hbar^2\epsilon^2E_i$)$^{1/2}$, with $E_i$ the Mn acceptor level, $m$*/m$_0$ the effective mass ratio determined from the effective Bohr radius $a^*$=$\hbar^2\epsilon$/$m^*e^2$, with $\epsilon$ the static dielectric constant. Following Kojima $et$ al.~\cite{Kojima2007}, the factor $E_{\mathrm{eff}}$/$E_0$ is set equal to 1. 

In order to adapt this model to the case of Ga$_{1-x}$Mn$_x$As, and again following Kojima $et$ al.~\cite{Kojima2007}, we add background and disorder effects. To take disorder into account we use a phenomenological Gaussian smearing with broadening factor $\mu$, resulting in the modified cross-section,

\begin{equation}
{ S_{BR}(\hbar \omega, E_i) = \frac{1}{\sqrt{2\pi \mu}} \int_{-\infty}^{\infty}{e^{-(E'-E_i)^2/2\mu^2} S_{BR}(\hbar \omega, E')dE'}.       }
\label{smear}
\end{equation}

\noindent The absorption spectrum is then given by,

\begin{equation}
{\alpha(\hbar \omega, E_i)=NS_{BR}(\hbar \omega, E_i)  + \alpha_B  },
\end{equation}

\noindent where $\alpha_B$ is the background absorption assumed to be constant in this model, and $N$ is the density of neutral Mn acceptors available to accept an electron from the VB to the Mn impurity level. Finally, the $\sigma_1(\omega)$ spectrum is calculated by $\sigma_1(\omega)$=$n(\hbar \omega)\alpha(\hbar \omega, E_i)$/(4$\pi\frac{1}{\lambda}$). The curve shown in the manuscript uses $\alpha_B$=14500 cm$^{-1}$, an acceptor density of $N$=10$^{20}$ cm$^{-3}$, and a broadening factor of $\mu$=40 meV taken from time-resolved spectrum in Ref.~\cite{Kojima2007}.

\subsection{low-$\omega$ analysis}

An agreement is expected between the low-$\omega$ $\sigma_1(\omega)$ extracted from optics and dc values provided there are not other strong electronic exitations between 40 cm$^{-1}$ and dc. While the gradient nature of the samples in this study complicates transport measuremensts, we note the dc conductivity for the Ga$_{0.84}$Mn$_{16}$As sample in this study are reported at the location of minumum compensation in Ref.~\cite{Mack2008}, and show excellent agreement with the $\sigma_1$(40 cm$^{-1}$) value extracted from our optical probe.

Interestingly, the full-width at half-max of the Drude peak, quantifying the scattering rate within the Drude model, is not enhanced in the $x$=0.009 Be-doped sample with respect to $x$=3.2$\times$10$^{-4}$ Be-doped. Recalling that only the former was grown in low-temperature conditions points towards two possible scenarios: either the disorder in samples grown under low-temperature conditions is similar to lower-doped conventionally grown films; or the increase in disorder is effectively screened out by excess charges~\cite{Jungwirth2007}. Note, neither of these scenarios is necessarily specific to the case of Be-doped GaAs. Furthermore, because $m_{\mathrm{opt}}$ is related to the band dispersion, the similar values found in the two metallic Be-doped samples suggests that whatever disorder is present, it is not significantly altering the nature of states in the vicinity of the Fermi level ($E_F$). Difficulties in accurately determining $p$ in magnetic systems, and in decoupling delocalized Drude-like transport from interband transition contributions limit similar analysis of the scattering rate and $m_{\mathrm{opt}}$ in Ga$_{1-x}$Mn$_x$As. However, reasonable estimates reveal the effective mass of carriers in Ga$_{1-x}$Mn$_x$As to be significatly larger than that of the metallic Be-doped samples. Detailed quantitave analysis of Ga$_{1-x}$Mn$_x$As effective masses in IR data can be found in Ref.~\cite{Singley2002, Burch2006}.




\begin{thebibliography}{45}
\expandafter\ifx\csname natexlab\endcsname\relax\def\natexlab#1{#1}\fi
\expandafter\ifx\csname bibnamefont\endcsname\relax
  \def\bibnamefont#1{#1}\fi
\expandafter\ifx\csname bibfnamefont\endcsname\relax
  \def\bibfnamefont#1{#1}\fi
\expandafter\ifx\csname citenamefont\endcsname\relax
  \def\citenamefont#1{#1}\fi
\expandafter\ifx\csname url\endcsname\relax
  \def\url#1{\texttt{#1}}\fi
\expandafter\ifx\csname urlprefix\endcsname\relax\def\urlprefix{URL }\fi

\bibitem[{\citenamefont{Coey et~al.}(1999)\citenamefont{Coey, Viret, and {Von
  Molnar}}}]{Coey1999}
\bibinfo{author}{\bibfnamefont{J.~M.~D.} \bibnamefont{Coey}} \bibnamefont{et~al.},
  \bibinfo{journal}{Adv. Phys.} \textbf{\bibinfo{volume}{48}},
  \bibinfo{pages}{167} (\bibinfo{year}{1999}).

\bibitem{Sato2010}
\bibinfo{author}{\bibfnamefont{K.}~\bibnamefont{Sato}} \bibnamefont{et~al.},
  \bibinfo{journal}{Rev. Mod. Phys.} \textbf{\bibinfo{volume}{82}},
  \bibinfo{pages}{1633} (\bibinfo{year}{2010}).

\bibitem[{\citenamefont{Myers et~al.}(2006)\citenamefont{Myers, Sheu, Jackson,
  Gossard, Schiffer, Samarth, and Awschalom}}]{Myers2006}
\bibinfo{author}{\bibfnamefont{R.~C.} \bibnamefont{Myers}} \bibnamefont{et~al.},
  \bibinfo{journal}{Phys. Rev. B} \textbf{\bibinfo{volume}{74}},
  \bibinfo{pages}{155203} (\bibinfo{year}{2006}).

\bibitem[{\citenamefont{Mack et~al.}(2008)\citenamefont{Mack, Myers, Heron,
  Gossard, and Awschalom}}]{Mack2008}
\bibinfo{author}{\bibfnamefont{S.}~\bibnamefont{Mack}} \bibnamefont{et~al.}, \bibinfo{journal}{Appl. Phys. Lett.}
  \textbf{\bibinfo{volume}{92}}, \bibinfo{pages}{192502}
  (\bibinfo{year}{2008}).
  
  \bibitem[{\citenamefont{Nagai and Nagasaka}(2006)}]{Nagai2005}
\bibinfo{author}{\bibfnamefont{Y.}~\bibnamefont{Nagai}} \bibnamefont{and}
  \bibinfo{author}{\bibfnamefont{K.}~\bibnamefont{Nagasaka}},
  \bibinfo{journal}{Infrared Phys. Techno.} \textbf{\bibinfo{volume}{48}},
  \bibinfo{pages}{1} (\bibinfo{year}{2006}).

  

\bibitem[{\citenamefont{Sheu et~al.}(2007)\citenamefont{Sheu, Myers, Tang,
  Samarth, Awschalom, Schiffer, and Flatte´}}]{Sheu2007}
\bibinfo{author}{\bibfnamefont{B.~L.} \bibnamefont{Sheu}} \bibnamefont{et~al.},
  \bibinfo{journal}{Phys. Rev. Lett.} \textbf{\bibinfo{volume}{99}},
  \bibinfo{pages}{227205} (\bibinfo{year}{2007}).

\bibitem[{\citenamefont{Moriya and Munekata}(2003)}]{Moriya2003}
\bibinfo{author}{\bibfnamefont{R.}~\bibnamefont{Moriya}} \bibnamefont{and}
  \bibinfo{author}{\bibfnamefont{H.}~\bibnamefont{Munekata}},
  \bibinfo{journal}{J. Appl. Phys.} \textbf{\bibinfo{volume}{93}},
  \bibinfo{pages}{4603} (\bibinfo{year}{2003}).

\bibitem[{\citenamefont{Jungwirth et~al.}(2007)\citenamefont{Jungwirth, Sinova,
  MacDonald, Gallagher, Nov\'{a}k, Edmonds, Rushforth, Campion, Foxon, Eaves
  et~al.}}]{Jungwirth2007}
\bibinfo{author}{\bibfnamefont{T.}~\bibnamefont{Jungwirth}}
  \bibnamefont{et~al.}, \bibinfo{journal}{Phys. Rev. B}
  \textbf{\bibinfo{volume}{76}}, \bibinfo{pages}{125206} (\bibinfo{year}{2007}).
  

\bibitem[{\citenamefont{Okabayashi et~al.}(2001)\citenamefont{Okabayashi,
  Kimura, Rader, Mizokawa, Fujimori, Hayashi, and Tanaka}}]{Okabayashi2001}
\bibinfo{author}{\bibfnamefont{J.}~\bibnamefont{Okabayashi}} \bibnamefont{et~al.},
  \bibinfo{journal}{Phys. Rev. B} \textbf{\bibinfo{volume}{64}},
  \bibinfo{pages}{125304} (\bibinfo{year}{2001}).

\bibitem[{\citenamefont{Singley et~al.}(2002)\citenamefont{Singley, Kawakami,
  Awschalom, and Basov}}]{Singley2002}
\bibinfo{author}{\bibfnamefont{E.~J.} \bibnamefont{Singley}} \bibnamefont{et~al.},
  \bibinfo{journal}{Phys. Rev. Lett.} \textbf{\bibinfo{volume}{89}},
  \bibinfo{pages}{097203} (\bibinfo{year}{2002}).
  
  \bibitem[{\citenamefont{Burch et~al.}(2006)\citenamefont{Burch, Shrekenhamer,
  Singley, Stephens, Sheu, Kawakami, Schiffer, Samarth, Awschalom, and
  Basov}}]{Burch2006}
\bibinfo{author}{\bibfnamefont{K.~S.} \bibnamefont{Burch}} \bibnamefont{et~al.},
  \bibinfo{journal}{Phys. Rev. Lett.} \textbf{\bibinfo{volume}{97}},
  \bibinfo{pages}{087208} (\bibinfo{year}{2006}).
  
    \bibitem[{\citenamefont{Kojima et~al.}(2007)\citenamefont{Kojima, H\'{e}roux,
  Shimano, Hashimoto, Katsumoto, Iye, and Kuwata-Gonokami}}]{Kojima2007}
\bibinfo{author}{\bibfnamefont{E.}~\bibnamefont{Kojima}} \bibnamefont{et~al.},
  \bibinfo{journal}{Phys. Rev. B} \textbf{\bibinfo{volume}{76}},
  \bibinfo{pages}{195323} (\bibinfo{year}{2007}).


\bibitem[{\citenamefont{Ando et~al.}(2008)\citenamefont{Ando, Saito, Agarwal,
  Debnath, and Zayets}}]{Ando2008}
\bibinfo{author}{\bibfnamefont{K.}~\bibnamefont{Ando}} \bibnamefont{et~al.},
  \bibinfo{journal}{Phys. Rev. Lett.} \textbf{\bibinfo{volume}{100}},
  \bibinfo{pages}{067204} (\bibinfo{year}{2008}).

\bibitem[{\citenamefont{Ohya et~al.}(2010)\citenamefont{Ohya, Muneta, Hai, and
  Tanaka}}]{Ohya2010a}
\bibinfo{author}{\bibfnamefont{S.}~\bibnamefont{Ohya}} \bibnamefont{et~al.},
  \bibinfo{journal}{Phys. Rev. Lett.} \textbf{\bibinfo{volume}{104}},
  \bibinfo{pages}{167204} (\bibinfo{year}{2010}).

\bibitem[{\citenamefont{Mayer et~al.}(2010)\citenamefont{Mayer, Stone, Miller,
  Smith, Dubon, Haller, Yu, Walukiewicz, Liu, and Furdyna}}]{Mayer2010}
\bibinfo{author}{\bibfnamefont{M.~A.} \bibnamefont{Mayer}} \bibnamefont{et~al.},
  \bibinfo{journal}{Phys. Rev. B} \textbf{\bibinfo{volume}{81}},
  \bibinfo{pages}{045205} (\bibinfo{year}{2010}).
  
\bibitem[{\citenamefont{Sapega et~al.}(2010)\citenamefont{Sapega, Moreno, Ransteiner,
  Daweritz, and Ploog}}]{Sapega2005}
\bibinfo{author}{\bibfnamefont{V.~F.} \bibnamefont{Sapega}} \bibnamefont{et~al.},
  \bibinfo{journal}{Phys. Rev. Lett.} \textbf{\bibinfo{volume}{94}},
  \bibinfo{pages}{137401} (\bibinfo{year}{2005}).
  
  \bibitem[{\citenamefont{Rokhinson et~al.}(2007)\citenamefont{Rokhinson, Lyanda-Geller, Ge,
  Shen, Lie, Dobrowolska, and Furdyna}}]{Rokhinson2007}
\bibinfo{author}{\bibfnamefont{L.~P.} \bibnamefont{Rokhinson}} \bibnamefont{et~al.},
  \bibinfo{journal}{Phys. Rev. B} \textbf{\bibinfo{volume}{76}},
  \bibinfo{pages}{161201(R)} (\bibinfo{year}{2007}).





\bibitem[{\citenamefont{Okimoto et~al.}(1995)\citenamefont{Okimoto, Katsufuji,
  Ishikawa, Urushibara, Arima, and Tokura}}]{Okimoto1995}
\bibinfo{author}{\bibfnamefont{Y.}~\bibnamefont{Okimoto}} \bibnamefont{et~al.},
  \bibinfo{journal}{Phys. Rev. Lett.} \textbf{\bibinfo{volume}{75}},
  \bibinfo{pages}{109} (\bibinfo{year}{1995}).

\bibitem[{\citenamefont{Hirakawa}(2001)}]{Hirakawa2001}
\bibinfo{author}{\bibfnamefont{K.}~\bibnamefont{Hirakawa}},
  \bibinfo{journal}{Physica E}
  \textbf{\bibinfo{volume}{10}}, \bibinfo{pages}{215} (\bibinfo{year}{2001}).
  
  \bibitem[{\citenamefont{Basov}(2011)}]{Basov2011}
\bibinfo{author}{\bibfnamefont{D.~N.}~\bibnamefont{Basov}},
  \bibinfo{journal}{Rev. Mod. Phys.}
  \textbf{\bibinfo{volume}{83}}, \bibinfo{pages}{471} (\bibinfo{year}{2011}).

  
\bibitem{Romero1990}
\bibinfo{author}{\bibfnamefont{D.}~\bibnamefont{Romero}} \bibnamefont{et~al.},
  \bibinfo{journal}{Phys. Rev. B} \textbf{\bibinfo{volume}{42}},
  \bibinfo{pages}{3179} (\bibinfo{year}{1990}).
  
\bibitem{Gaymann1995}
\bibinfo{author}{\bibfnamefont{A.}~\bibnamefont{Gaymann}} \bibnamefont{et~al.},
  \bibinfo{journal}{Phys. Rev. B} \textbf{\bibinfo{volume}{52}},
  \bibinfo{pages}{16486} (\bibinfo{year}{1995}).    
  
\bibitem{cutoff} Choice in cut-off excludes contributions from excitations into the GaAs conduction band, and facilitates direct comparison with other studies in the literature [11, 23].

\bibitem[{\citenamefont{Sinova et~al.}(2002)\citenamefont{Sinova, Jungwirth,
  Yang, Ku\v{c}era, and MacDonald}}]{Sinova2002}
\bibinfo{author}{\bibfnamefont{J.}~\bibnamefont{Sinova}} \bibnamefont{et~al.}, \bibinfo{journal}{Phys. Rev. B}
  \textbf{\bibinfo{volume}{66}}, \bibinfo{pages}{041202} (\bibinfo{year}{2002}).
  
  \bibitem[{\citenamefont{Songprakob et~al.}(2002)\citenamefont{Songprakob,
  Zallen, Tsu, and Liu}}]{Songprakob2002}
\bibinfo{author}{\bibfnamefont{W.}~\bibnamefont{Songprakob}} \bibnamefont{et~al.},
  \bibinfo{journal}{J. Appl. Phys.} \textbf{\bibinfo{volume}{91}},
  \bibinfo{pages}{171} (\bibinfo{year}{2002}).

\bibitem[{\citenamefont{Braunstein and Kane}(1962)}]{Braunstein1962}
\bibinfo{author}{\bibfnamefont{R.}~\bibnamefont{Braunstein}} \bibnamefont{and}
  \bibinfo{author}{\bibfnamefont{E.~O.} \bibnamefont{Kane}},
  \bibinfo{journal}{J. Phys. Chem. Solids} \textbf{\bibinfo{volume}{23}},
  \bibinfo{pages}{1423} (\bibinfo{year}{1962}).
  
   \bibitem[{\citenamefont{Moca et~al.}(2009)\citenamefont{Moca, Zar\'{a}nd, and
  Berciu}}]{Moca2009a}
\bibinfo{author}{\bibfnamefont{C.~P.} \bibnamefont{Moca}} \bibnamefont{et~al.},
  \bibinfo{journal}{Phys. Rev. B} \textbf{\bibinfo{volume}{80}},
  \bibinfo{pages}{165202} (\bibinfo{year}{2009}).
  
  \bibitem[{\citenamefont{Bouzerar and
  Bouzerar}(2010{\natexlab{b}})}]{Bouzerar2010a}
\bibinfo{author}{\bibfnamefont{G.}~\bibnamefont{Bouzerar}} \bibnamefont{and}
  \bibinfo{author}{\bibfnamefont{R.}~\bibnamefont{Bouzerar}},
  \bibinfo{journal}{New J. Phys.}  \textbf{\bibinfo{volume}{13}}, \bibinfo{pages}{023002}
  (\bibinfo{year}{2011}{\natexlab{b}}).
  
    \bibitem[{\citenamefont{Masek et~al.}(2010)\citenamefont{Jungwirth, Sinova,
  Ma\v{s}ek, Ku\v{c}era, and MacDonald}}]{Masek2010}
\bibinfo{author}{\bibfnamefont{J.}~\bibnamefont{Ma\v{s}ek}} \bibnamefont{et~al.},
  \bibinfo{journal}{Phys. Rev. Lett.} \textbf{\bibinfo{volume}{105}},
  \bibinfo{pages}{227202} (\bibinfo{year}{2010}).

\bibitem[{\citenamefont{Yu et~al.}(2002)\citenamefont{Yu, Walukiewicz,
  Wojtowicz, Kuryliszyn, Liu, Sasaki, and Furdyna}}]{Yu2002}
\bibinfo{author}{\bibfnamefont{K.~M.} \bibnamefont{Yu}} \bibnamefont{et~al.},
  \bibinfo{journal}{Phys. Rev. B} \textbf{\bibinfo{volume}{65}},
  \bibinfo{pages}{201303} (\bibinfo{year}{2002}).

\bibitem[{\citenamefont{Dunsiger et~al.}(2010)\citenamefont{Dunsiger, Carlo,
  Goko, Nieuwenhuys, Prokscha, Suter, Morenzoni, Chiba, Nishitani, Marsukura
  et~al.}}]{Dunsiger2010}
\bibinfo{author}{\bibfnamefont{S.~R.} \bibnamefont{Dunsiger}} \bibnamefont{et~al.}, \bibinfo{journal}{Nature Mater.}
  \textbf{\bibinfo{volume}{9}}, \bibinfo{pages}{299} (\bibinfo{year}{2010}).

\bibitem[{\citenamefont{Richardella et~al.}(2010)\citenamefont{Richardella,
  Roushan, Mack, Zhou, Huse, Awschalom, and Yazdani}}]{Richardella2010a}
\bibinfo{author}{\bibfnamefont{A.}~\bibnamefont{Richardella}} \bibnamefont{et~al.},
  \bibinfo{journal}{Science} \textbf{\bibinfo{volume}{327}},
  \bibinfo{pages}{665} (\bibinfo{year}{2010}).

\bibitem[{\citenamefont{Sapega et~al.}(2009)\citenamefont{Sapega,
  Sablina, Panaiotti, Averkiev, and Ploog}}]{Sapega2009}
\bibinfo{author}{\bibfnamefont{V.~F.}~\bibnamefont{Sapega}} \bibnamefont{et~al.},
  \bibinfo{journal}{Phys. Rev. B} \textbf{\bibinfo{volume}{80}},
  \bibinfo{pages}{041202(R)} (\bibinfo{year}{2009}).

\bibitem[{\citenamefont{Jungwirth et~al.}(2006)\citenamefont{Jungwirth, Sinova,
  Ma\v{s}ek, Ku\v{c}era, and MacDonald}}]{Jungwirth2006}
\bibinfo{author}{\bibfnamefont{T.}~\bibnamefont{Jungwirth}} \bibnamefont{et~al.}, \bibinfo{journal}{Rev. Mod. Phys.}
  \textbf{\bibinfo{volume}{78}}, \bibinfo{pages}{809} (\bibinfo{year}{2006}).

\bibitem[{\citenamefont{Dietl et~al.}(2000)\citenamefont{Dietl, Ohno,
  Marsukura, Cibert, and Ferrand}}]{Dietl2000}
\bibinfo{author}{\bibfnamefont{T.}~\bibnamefont{Dietl}} \bibnamefont{et~al.},
  \bibinfo{journal}{Science} \textbf{\bibinfo{volume}{287}},
  \bibinfo{pages}{1019} (\bibinfo{year}{2000}).
  
  
  \bibitem[{\citenamefont{Bouzerar and
  Bouzerar}(2010{\natexlab{a}})}]{Bouzerar2010b}
\bibinfo{author}{\bibfnamefont{R.}~\bibnamefont{Bouzerar}} \bibnamefont{and}
  \bibinfo{author}{\bibfnamefont{G.}~\bibnamefont{Bouzerar}},
  \bibinfo{journal}{EPL-Europhys. Lett.} \textbf{\bibinfo{volume}{92}},
  \bibinfo{pages}{47006} (\bibinfo{year}{2010}{\natexlab{a}}).
  
  

\bibitem[{\citenamefont{Nili et~al.}(2010)\citenamefont{Nili, Yu, Moreno,
  Browne, and Jarrell}}]{Nili2010a}
\bibinfo{author}{\bibfnamefont{A.-M.} \bibnamefont{Nili}} \bibnamefont{et~al.},
  \bibinfo{journal}{CondMat arxiv:1007.4609V1}  (\bibinfo{year}{2010}).

\bibitem[{\citenamefont{Chattopadhyay et~al.}(2001)\citenamefont{Chattopadhyay,
  {Das Sarma}, and Millis}}]{Chattopadhyay2001}
\bibinfo{author}{\bibfnamefont{A.}~\bibnamefont{Chattopadhyay}} \bibnamefont{et~al.},
  \bibinfo{journal}{Phys. Rev. Lett.} \textbf{\bibinfo{volume}{87}},
  \bibinfo{pages}{227202} (\bibinfo{year}{2001}).
  



\bibitem[{\citenamefont{Edmonds et~al.}(2004)\citenamefont{Edmonds,
  Boguslawski, Wang, Campion, Novikov, Farley, Gallagher, Foxon, Sawicki,
  Dietl et~al.}}]{Edmonds2004}
\bibinfo{author}{\bibfnamefont{K.}~\bibnamefont{Edmonds}},
  \bibinfo{author}{\bibfnamefont{P.}~\bibnamefont{Boguslawski}},
  \bibinfo{author}{\bibfnamefont{K.}~\bibnamefont{Wang}},
  \bibinfo{author}{\bibfnamefont{R.}~\bibnamefont{Campion}},
  \bibinfo{author}{\bibfnamefont{S.}~\bibnamefont{Novikov}},
  \bibinfo{author}{\bibfnamefont{N.}~\bibnamefont{Farley}},
  \bibinfo{author}{\bibfnamefont{B.}~\bibnamefont{Gallagher}},
  \bibinfo{author}{\bibfnamefont{C.}~\bibnamefont{Foxon}},
  \bibinfo{author}{\bibfnamefont{M.}~\bibnamefont{Sawicki}},
  \bibinfo{author}{\bibfnamefont{T.}~\bibnamefont{Dietl}},
  \bibnamefont{et~al.}, \bibinfo{journal}{Phys. Rev. Lett.}
  \textbf{\bibinfo{volume}{92}}, \bibinfo{pages}{037201} (\bibinfo{year}{2004}).

\bibitem[{\citenamefont{Liu et~al.}(1995)\citenamefont{Liu, Prasad, Nishio,
  Weber, Liliental-Weber, and Walukiewicz}}]{Liu1995}
\bibinfo{author}{\bibfnamefont{X.}~\bibnamefont{Liu}},
  \bibinfo{author}{\bibfnamefont{A.}~\bibnamefont{Prasad}},
  \bibinfo{author}{\bibfnamefont{J.}~\bibnamefont{Nishio}},
  \bibinfo{author}{\bibfnamefont{E.}~\bibnamefont{Weber}},
  \bibinfo{author}{\bibfnamefont{Z.}~\bibnamefont{Liliental-Weber}},
  \bibnamefont{and}
  \bibinfo{author}{\bibfnamefont{W.}~\bibnamefont{Walukiewicz}},
  \bibinfo{journal}{Appl. Phys. Lett.} \textbf{\bibinfo{volume}{67}},
  \bibinfo{pages}{279} (\bibinfo{year}{1995}).

\bibitem[{\citenamefont{Missous}(1994)}]{Missous1994}
\bibinfo{author}{\bibfnamefont{M.}~\bibnamefont{Missous}}, \bibinfo{journal}{J.
  Appl. Phys.} \textbf{\bibinfo{volume}{75}}, \bibinfo{pages}{3396}
  (\bibinfo{year}{1994}).



\bibitem[{\citenamefont{Wooten}(1972)}]{Wooten1972}
\bibinfo{author}{\bibfnamefont{F.}~\bibnamefont{Wooten}},
  \emph{\bibinfo{title}{{Optical Properties of Solids}}}
  (\bibinfo{publisher}{Academic}, \bibinfo{address}{New York, London},
  \bibinfo{year}{1972}).


        
  \bibitem[{\citenamefont{Bebb}(1967)}]{Bebb1967}
\bibinfo{author}{\bibfnamefont{H.}~\bibnamefont{Bebb}},
  \bibinfo{journal}{J. Phys. Chem. Solids}
  \textbf{\bibinfo{volume}{28}}, \bibinfo{pages}{2087} (\bibinfo{year}{1967}).

\bibitem[{\citenamefont{Bebb}(1969)}]{Bebb1969}
\bibinfo{author}{\bibfnamefont{H.}~\bibnamefont{Bebb}},
  \bibinfo{journal}{Phys. Rev.} \textbf{\bibinfo{volume}{185}},
  \bibinfo{pages}{1116} (\bibinfo{year}{1969}).

\bibitem[{\citenamefont{Huber et~al.}(1987)\citenamefont{Huber, Perez, and
  Huber}}]{Huber1987}
\bibinfo{author}{\bibfnamefont{C.}~\bibnamefont{Huber}},
  \bibinfo{author}{\bibfnamefont{J.}~\bibnamefont{Perez}}, \bibnamefont{and}
  \bibinfo{author}{\bibfnamefont{T.}~\bibnamefont{Huber}},
  \bibinfo{journal}{Phys. Rev. B} \textbf{\bibinfo{volume}{36}},
  \bibinfo{pages}{5933} (\bibinfo{year}{1987}).

  
  
  
  
 

  
 

\end{thebibliography}
\end{document}